\documentclass[12pt,a4paper,twoside, openright]{report}
\usepackage[twoside,width=16cm,height=21cm,left=3.7cm,right=2.7cm]{geometry}
\usepackage[utf8]{inputenc}
\usepackage{polski}
\usepackage[english,polish]{babel}
\usepackage{amsmath}
\usepackage{lmodern}
\usepackage{graphicx}
\usepackage{placeins}
\usepackage{lscape}
\usepackage{afterpage}
\usepackage{emptypage}
\usepackage{geometry}
\usepackage{array}
\usepackage{caption}
\usepackage{subcaption}
\usepackage[square, numbers]{natbib}
\usepackage{sectsty}
\usepackage{float}
\usepackage{wrapfig}
\usepackage{chngcntr}
\usepackage[onehalfspacing]{setspace}
\author{Dominika Alfs}
\title{Test of~Lightguides for the \mbox{J-PET} Detector}

\chapterfont{\fontsize{14}{15}\selectfont}
\sectionfont{\fontsize{13}{15}\selectfont}
\subsectionfont{\fontsize{13}{15}\selectfont}

\begin{document}
\selectlanguage{polish}
\pagestyle{empty}
\begin{center}
\begin{large}
\textbf{Jagiellonian University in Kraków}\\
Faculty of Physics, Astronomy and Applied Computer Science\\
\vspace{112pt}
\textbf{{\normalsize Dominika Alfs}}\\
\end{large}
\vspace{60pt}
{\Large \textbf{Test of~Lightguides for the \mbox{J-PET} Detector}\\ \vspace{12pt}}
Bachelor Thesis\\
\vspace{12pt}
\end{center}
\vspace{150pt}
\begin{flushright}
Supervised by\\
dr Eryk Czerwiński\\
Institute of Physics\\
\foreignlanguage{polish}
Division of Nuclear Physics
\end{flushright}
\vspace{24pt}
\begin{center}
Kraków, 2015
\end{center}
\newpage
\null
\newpage
\null
\vspace{60pt}
\begin{center}
{\Large \textbf{Abstract}}
\end{center}
\vspace{24pt}
\selectlanguage{english}
The aim of this work was to test the impact of the lightguides insertion in the \mbox{\mbox{J-PET}} detection system and to choose the best of available solutions. At the two strip \mbox{J-PET} test system a series of measurements was performed with different combinations concerning the length of the lightguides (3 mm and 3 cm) and scintillators (3, 15 and 30 cm), shape of lightguides (trapezoidal and cylindrical, with and without a cut matching the scintillator size) and optical connection between elements (optical gel and glue). It was proven that the insertion of the thin lightguide does not spoil the time resolution and the light output. Additionally, the correlation between the time resolution and the light output was confirmed.
\newpage
\null
\newpage
\null
\vspace{60pt}
\begin{center}
{\Large \textbf{Abstrakt}}
\end{center}
\vspace{24pt}
\selectlanguage{polish}
Celem tej pracy było sprawdzenie wpływu dołączenia światłowodów do systemu detekcyjnego \mbox{\mbox{J-PET}} oraz wybór najlepszego z~dostępnych sposobów realizacji tego zagadnienia. Wykonano serię pomiarów z~wykorzystaniem dwupaskowego testowego systemu \mbox{\mbox{J-PET}}, uwzględniając różne kombinacje długości światłowodów (\mbox{3 mm} i~\mbox{3 cm}), ich kształu (trapezoidalne i~cylindryczne, gładkie oraz z~wcięciami pasującymi do rozmiaru scyntylatorów) oraz połączenia optycznego na granicy elementów (żel optyczny oraz klej optyczny). Udowodniono, że dodanie cienkiego światłowodu nie pogarsza czasowej zdolności rozdzielczej oraz ilości uzyskiwanego światła. Dodatkowo potwierdzono także korelację pomiędzy tymi wielkościami.
\newpage
\selectlanguage{english}
\null
\newpage
\pagenumbering{gobble}
\clearpage
\thispagestyle{empty}
\tableofcontents
\newpage
\chapter{Motivation}
\pagestyle{plain}
\pagenumbering{arabic}
Development of~medical imaging is experiencing a~lot of~interest nowadays. Among the most popular techniques there are for example computed tomography (CT) \cite{CT}, single photon emission computed tomography (SPECT) \cite{SPECT}, magnetic resonance imaging (MRI) \cite{MRI} and positron emission tomography (PET) \cite{PET}. New technologies has made it possible to~come across the increasing need of~creating more precise while less expensive systems.

One of~the most promising techniques for further improvement is PET. As~it is a~non-invasive technique, PET is an invaluable tool for a~variety of~medical purposes: monitoring functions of~certain organs, studying new drugs impact on organism and diagnostic processes in cardiology, oncology (including diagnosis of~early stages of~cancer) and even psychiatry.

Most common PET detectors are made of~crystal scintillators arranged in a~ring that surrounds the patient. However, this is an expensive solution: the size of~the area covered with crystals is strongly limited by the cost of~their production what not only decreases the detection efficiency but also makes it impossible to~perform eg. full body examination within one scan.

The solution of~these problems may become a~strip PET. The basis of~this idea is to~replace crystals by the polymer scintillators which are significantly cheaper and easier to~shape. This would not only decrease the price of~the device but also allow to~produce the detectors long enough to~cover all of~the patient's body. The prototype of~such a~device named Jagiellonian-PET (\mbox{J-PET}) is currently under construction at the Jagiellonian University, Cracow \cite{JPET, JPET2}.

The arrangement of~the scintillators and photomultipliers in \mbox{J-PET} is different than in PET. In case of~PET each crystal detector is connected to~a single photomultiplier, in \mbox{J-PET} there are two photomultipliers per single scintillator: one photomultiplier on each end of~the strip.

Taking under consideration the \mbox{J-PET} mechanics and geometry, it appears crucial to~provide a~stable connection between the photomultiplier and the scintillator. The solution discussed in this work is the insertion of lightguides between the scintillator and the photomultiplier.

The goal of~the performed investigation was to~test the impact of~the the lightguide usage in the two strip \mbox{J-PET} model and determine the most effective way of~their application. The criteria of~comparison were the time resolution and the light output of~the signals.

\FloatBarrier
\chapter{Introduction to~Positron Emission Tomography}
Unlike other medical imaging techniques, which provide the information about the location and shape of~the physical structures in the body, PET provides information about their metabolism.

In order to~be able to~observe any metabolic processes the radiopharmaceutical must be injected inside the patient's body. Radiopharmaceuticals are substances consisting of~two main parts: a~biologically active molecule and radionuclide (tracer) that undergoes a~$ \beta^{+} $ decay: $ 	_{Z} ^{A}N \longrightarrow  \hspace{3pt}_{Z-1}^{A}\hspace{-2pt}N' + e^{+} + \nu_{e} $, where $ N $ stands for a~nucleus of~mass number $ A~$ and atomic number $ Z $.

\begin{figure}
  \begin{center}
    \includegraphics[width=0.55\textwidth]{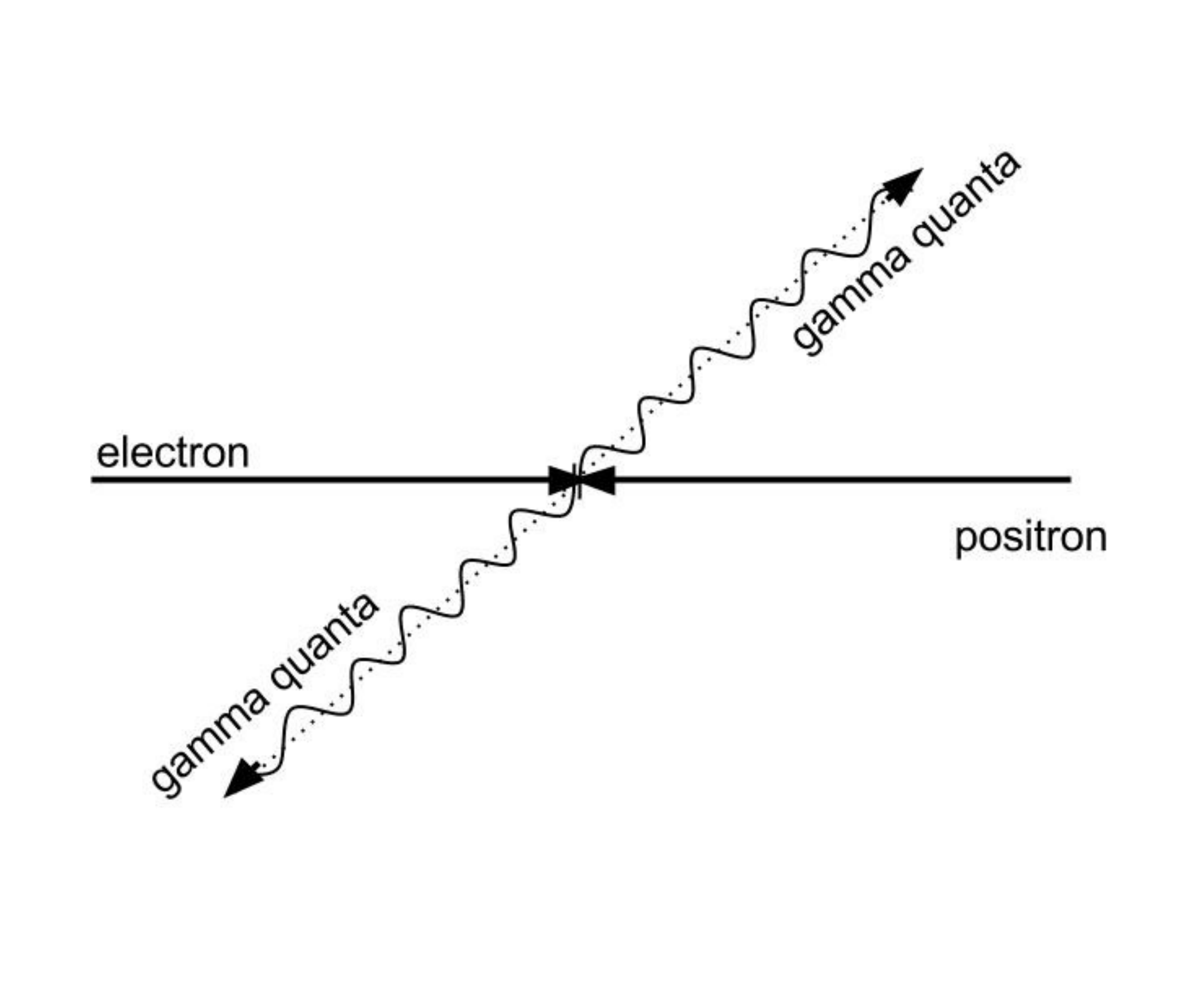}
  \end{center}
  \caption{\textit{A $ e^{+}e^{-} $ annihilation in the center of~mass frame.}}\label{fig:annihilation}
\end{figure}

\begin{figure}[!h]
\begin{center}
\includegraphics[width=\linewidth]{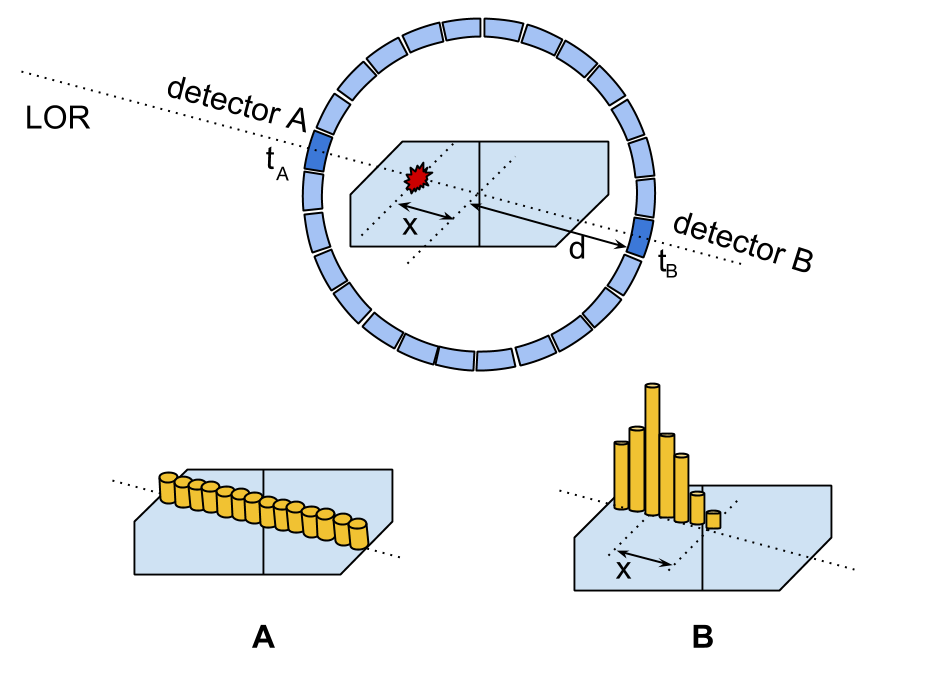}
\caption{\textit{TOF-PET idea. Annihilation occurs at~the distance $ x $ from the middle between two detectors. Two emitted gamma quanta hit two detectors which are at~the distance $ 2 d $ from each other: the first one after the time $ t_A $ and the second one after time $ t_B $. In case of~non-TOF reconstruction (A) the only information obtained is that the probability of~annihilation occurrence is uniform along the LOR for a~single event. In case of~TOF reconstruction (B) the difference between the detection times of~gamma quanta allows to~estimate the place of~the annihilation. $ x $ is the distance (along the LOR) form the center to~the point calculated as~$ \tfrac{1}{2}(t_{A} - t_{B})c$. The height of~bins of~the histograms in the picture corresponds to~the reconstructed probability of~annihilation occurrence in the given place.}}\label{TOF}
\end{center}
\end{figure}

Biologically active molecule is responsible for distribution of~the radiopharmaceutical inside the patient's body. It usually contains glucose - a~simple sugar which serves as~a~primary energy source in the human body metabolism processes. Within the illness development the cells perform intensive biochemical activity which corresponds to~increased energy consumption. For example actively growing cancer cells cause the radiopharmaceutical to~accumulate because of~the faster rate of~the glucose breakdown in comparison to~healthy tissues.
Positron emitted from the tracer travels a~short distance loosing its kinetic energy. The exact length depends on the isotope, eg. for a~positron of~energy 640~keV emitted form $^{18}$F the distance in tissue is 2.4 mm \cite{distance}.  Being almost at~rest a~positron can interact with an electron and an annihilation occurs. In over 99\% of~cases two gamma quanta of~energies equal 511~keV are emitted \cite{probability} (Figure~\ref{fig:annihilation}) and in rest of~cases there are more than two gamma quanta emitted (but the probability quickly decreases with their number). When gamma quantum manages to~reach the detector, it produces a~burst of~light in the scintillator which it hits. It is detected by the  photomultiplier connected with this scintillator. In order to~choose from all signals an interesting event, two signals need both to be: in the time coincidence and higher than some defined threshold. Single event provides enough information to~calculate one line of~response (LOR) which is the line connecting the hit positions of~gamma quanta in scintillators (Figure~\ref{TOF}). We claim that it is also the line along which the annihilation occurred. The place of~radiopharmaceutical concentration is reconstructed by crossing the set of~such lines. 

A novel idea is to~restrict the investigated LOR only to~the area close to~the point of~annihilation by the analysis of~differences in time of~flight (TOF) of~coincident photons. This approach is advantageous not only due to~the image reconstruction fastening. It also causes the noise reduction and improves the sharpness of~obtained images. There is a~brief explanation of~the TOF-PET method presented in the Figure~\ref{TOF}.

The description above is only to~explain the basic idea of~the PET image reconstruction. Actually, it is a~complicated process concerning measurement method and physical effects. More detailed description of the PET techniques can be found in \cite{imageReco1, imageReco2}.

\FloatBarrier
\chapter{Jagiellonian-PET}
\vspace{-18pt}
As all of~the PET detectors for medical applications are based on the crystal scintillators, the idea of~usage the organic ones is an entirely new approach \cite{JPET}. Until now, they were not considered as~an effective solution because of~a low density and a~small atomic number in comparison with the inorganic ones. Low density implies that some of~gamma quanta do not interact within the volume of~the scintillator and the detection becomes less effective. Small atomic number is responsible for a~negligible probability of~deposited energy transfer through the photoelectric effect. Nevertheless, the new PET has been designed in a~way to~reduce these drawbacks and take the advantage of~other properties of~organic scintillators~\cite{lukasz, anna}.

\begin{figure}[!ht]
\begin{center}
\includegraphics[scale=0.4]{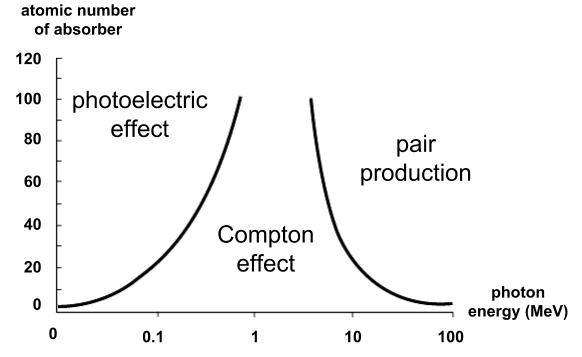}
\caption{\textit{Dominant effects of~photons interaction with matter. Each gamma quantum from $e^+e^-$ annihilation process is 0.511~MeV (or lower due to the scattering). In this range of~energies Compton effect is a~dominant one for small atomic numbers and photoelectric effect for greater ones. This is a~simple explanation of~effects present in scintillators. Polymer scintillators are mostly composed of~carbon (atomic number $ A_C =6 $) and hydrogen ($ A_H = 1 $) while crystal scintillators consist elements like cesium (in $ CsI(Tl) $, $ A_{Cs} = 55 $), thalium (in $ NaI(Tl) $, $ A_{Tl} = 81 $), lead (in $ Pb_4WO_4 $, $ A_{Pb} = 82 $) or bismuth ($ Bi_4Ge_3O_{12} $, $ A_{Bi} = 83 $). Data adapted from \cite{dominantEffect}.}}\label{dominantEffects}
\end{center}
\end{figure}

\begin{figure}[!hb]
\centering
\includegraphics[width=0.6\linewidth]{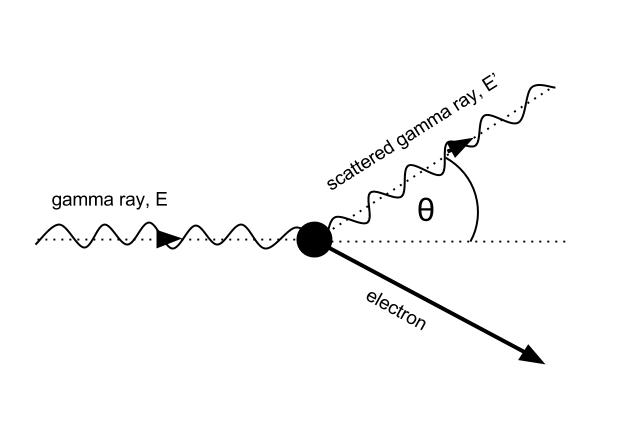}
\caption{\textit{Compton effect is an elastic scattering of~an incident photon of~the energy E on a~free or loosely bound electron of~mass $ m_e $. The energy $ E' $ of~the scattered photon depends on its primary energy and the scattering angle $ \Theta $: $ E' = E(1 + E/(m_e c^2(1 - \cos \Theta)))^{-1} $. }}\label{compton}
\end{figure}

\begin{figure}[!hb]
\begin{center}
\includegraphics[width=0.82\linewidth]{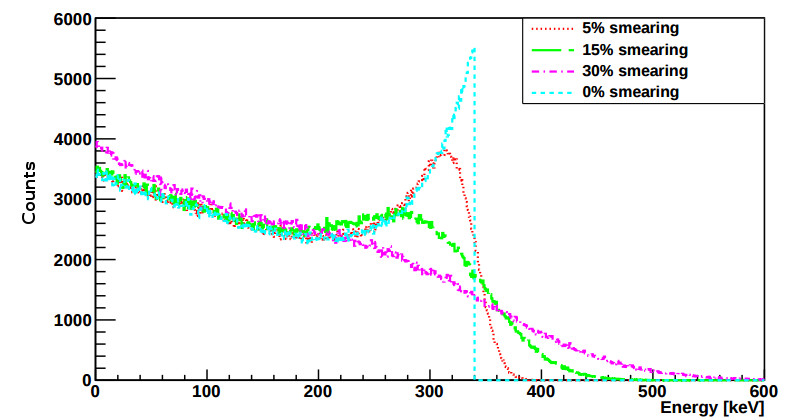}
\caption{\textit{Distribution of~kinetic energy of~electron assuming different energy resolution. Simulations performed for the incident gamma quanta energy of~511~keV. Figure adapted from \cite{szymon}.}}\label{kineticEnergy}
\end{center}
\end{figure}

\begin{figure}[!hb]
\begin{center}
\includegraphics[width=0.82\linewidth]{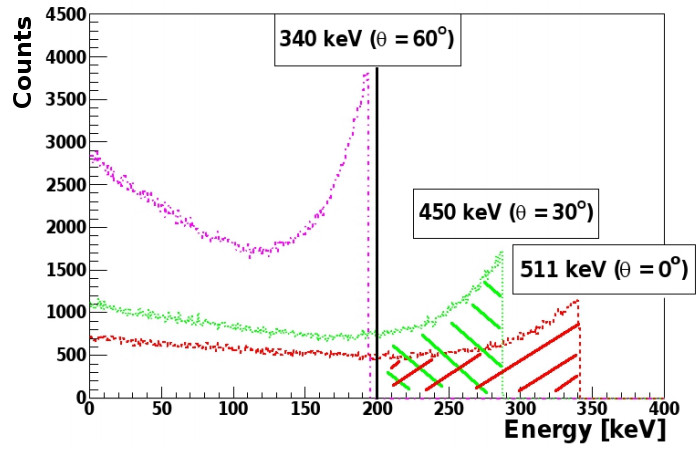}
\caption{\textit{Simulated distribution of energy of the electrons gained via Compton effect for three different energies of an incident $ \gamma $ quanta. The vertical black line corresponds to the threshold. Only electrons with kinetic energy higher than this threshold are accepted in further analysis. Figure adapted from \cite{szymon}.}}\label{scatteringAngleEnergy}
\end{center}
\end{figure}

Even though in the plastic scintillator the energy cannot be transferred by the gamma quanta in the photoelectric effect it is still possible through the Compton effect (Figures~\ref{dominantEffects}, \ref{compton}). In currently used PET systems events scattered more than 60 degrees are rejected by the energy cut around the photoelectric peak. However it is also possible to~apply the energy cut on the Compton spectra registered by the organic scintillating strips. Theoretical Compton spectrum for the annihilation quanta with an effect coming from the detection resolution is shown in the Figure~\ref{kineticEnergy}, while the change of~the scattered gamma quanta as~a~function of~the scattering angle is presented in the Figure~\ref{scatteringAngleEnergy}. Apart of mentioned earlier drawbacks of lower density of plastic scintillators one can see that is has also a~positive outcome. In the crystal scintillators the light is significantly more attenuated than in the polymer ones (the light attenuation length in organic scintillators is of~about 2~m, while in inorganic ones of~about 2~cm \cite{attenuationLengths}). This feature connected with the simplicity of~shaping polymers gives an opportunity to~produce long scintillating strips which provide the light output sufficient for further analysis.

\begin{figure}
\begin{center}
\includegraphics[width=0.5\linewidth]{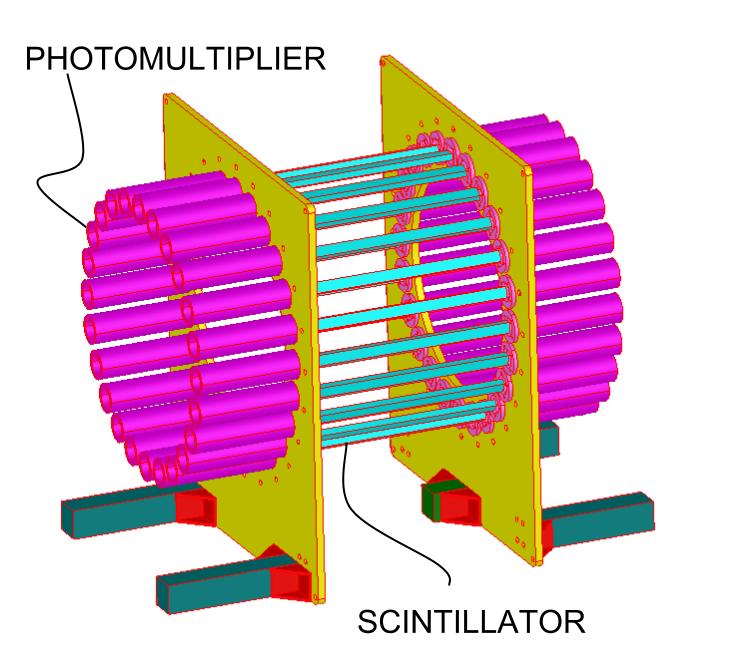}
\caption{\textit{Schematic view of the full \mbox{J-PET} detection system. Two photomultipliers are connected to~the single scintillator forming a~cylindrical chamber. Figure adapted from \cite{heczko}.}}\label{JPETarrangement}
\end{center}
\end{figure}

The diagnostic chamber of~the \mbox{J-PET} consists of~a~few cylindrical layer, made of~long, organic scintillators with photomultipliers on both ends of~each strip (Figure~\ref{JPETarrangement}).  The increased length of~the device will provide a~solid angle greater than in conventional PET resulting in improvement of~gamma quanta acceptance. Furthermore, this will let perform  a~simultaneous imaging of~the biological processes inside the whole body. Application of multiple layers of~scintillators would increase the probability of~gamma quanta interaction with the detector volume. Distinction between the layers decreases the problem of~an unknown depth of~interaction in the material and thus the usage of~multiple layers instead the thicker one is reasonable.

Last but not least, the \mbox{J-PET} will not only provide data of~better precision than possible to~obtain nowadays (current PET scanners achieve the time resolution of~about 600 ps \cite{attenuationLengths}, at~present the \mbox{J-PET} time resolution in 125 ps \cite{JPETaccuracy}), but application of~the organic materials and the enlargement of~the diagnostic chamber without the necessity of~increasing the number of~photomultipliers (in comparison to~standard devices) will decrease the cost of~the production making this imaging technique more accessible.

\vspace{32pt}

\FloatBarrier
\chapter{Measurements of~Lightguides Properties}
\FloatBarrier
\section{Experimental setup}
\FloatBarrier
In order to~obtain reliable results from any measurement with two strip \mbox{J-PET} prototype it is necessary to~pay attention on the connection between the scintillator and photomultipliers. First of~all, the lost of~light at~the surface should be as~small as~possible. This connection has also to~be stable in time, symmetrical on both ends of~the scintillator and easy to~reconnect after once dismounted.

In the current setup this connection is provided by the thin layer of~the optical gel between the scintillator and the photomultipliers. This requires some effort to~mount the elements correctly. In case of~more complex systems consisted of~greater number of~elements this solution would be ineffective and difficult to~perform in a~proper way.

The proposed idea is the insertion of~the lightguide between the photomultiplier and the scintillator. Lightguides are easy to~shape therefore it is possible to~prepare them so that one end fit exactly the scintillator cross section and another provides easy connection with the photomultiplier. One of~possibilities for better stabilization is gluing the lightguide and the scintillator with the optical glue. 

\begin{figure}[!ht]
\begin{center}
\begin{subfigure}[c]{0.55\textwidth}
\includegraphics[width=\linewidth]{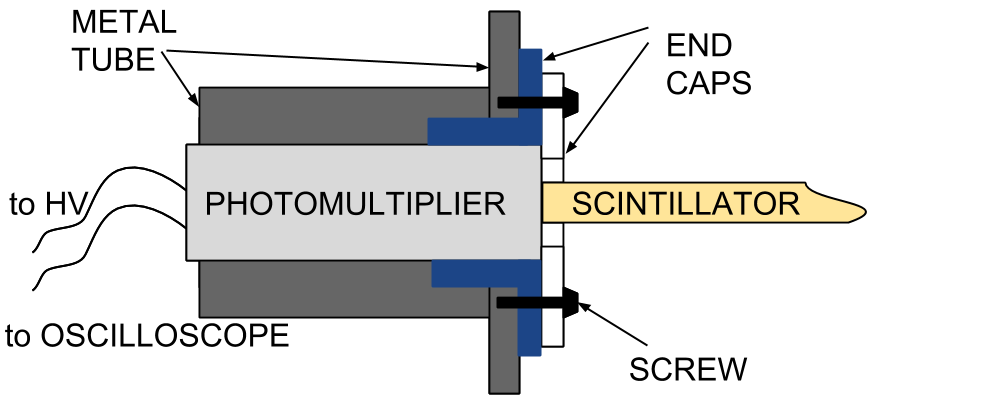}
\caption{}
\end{subfigure}
\begin{subfigure}[c]{0.55\textwidth}
\includegraphics[width=\linewidth]{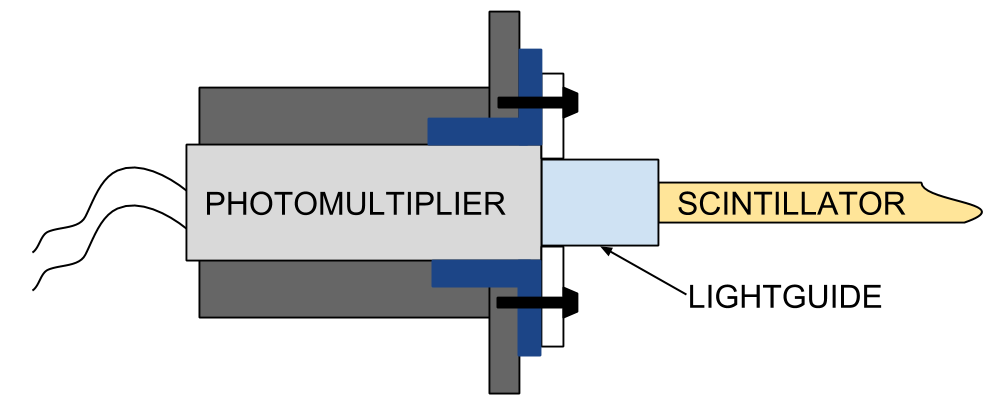}
\caption{}
\end{subfigure}
\caption{\textit{a) Current arrangement of~elements in the two strip \mbox{J-PET} system. The photomultiplier is  located inside metal tube. Between the scintillator and the photomultiplier there is a~thin layer of~the optical gel. End caps are inserted in order to~stabilize the connection. b) The idea of~lightguide insertion between the photomultiplier and the scintillator.}}\label{twoVer}
\end{center}
\end{figure}

As usage of~the material of~very good light guiding properties is a~natural idea (Figure~\ref{twoVer}), there still remains the question about the most effective shape of~the lightguides and the substance used at~the connection between them and the scintillator (optical gel BC630 or optical glue BC600).

\begin{figure}[!ht]
\begin{center}
\includegraphics[width=0.4\textwidth]{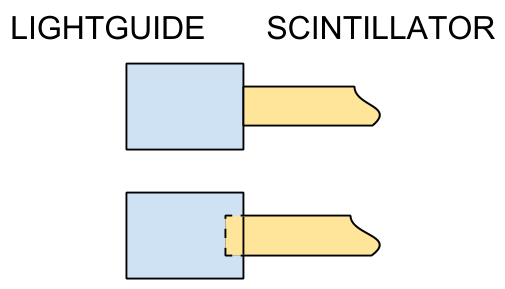}
\end{center}
\caption{\textit{Possible ways of~shaping the lightguide on the scintillator end.}}\label{shapesNew}
\end{figure}

Taking under consideration the geometry of~the elements, there were two shapes chosen for further investigation: a~cylindrical and a~trapezoidal one. After the measurements with each of~this shapes there the rows were cut in the lightguides matching exactly the scintillator shape (Figure~\ref{shapesNew}). The last shape that was checked were the trapezoidal lightguides with the cuts for the scintillator shortened up to~3~mm (Figure~\ref{photos}). The scheme and the dimensions of~the scintillator and the lightguides used are gathered in the Table \ref{dimensions}. Optical gel and glue were also tested.

\vspace{15pt}

\begin{figure}[!hb]
\begin{center}
\includegraphics[width=0.7\textwidth]{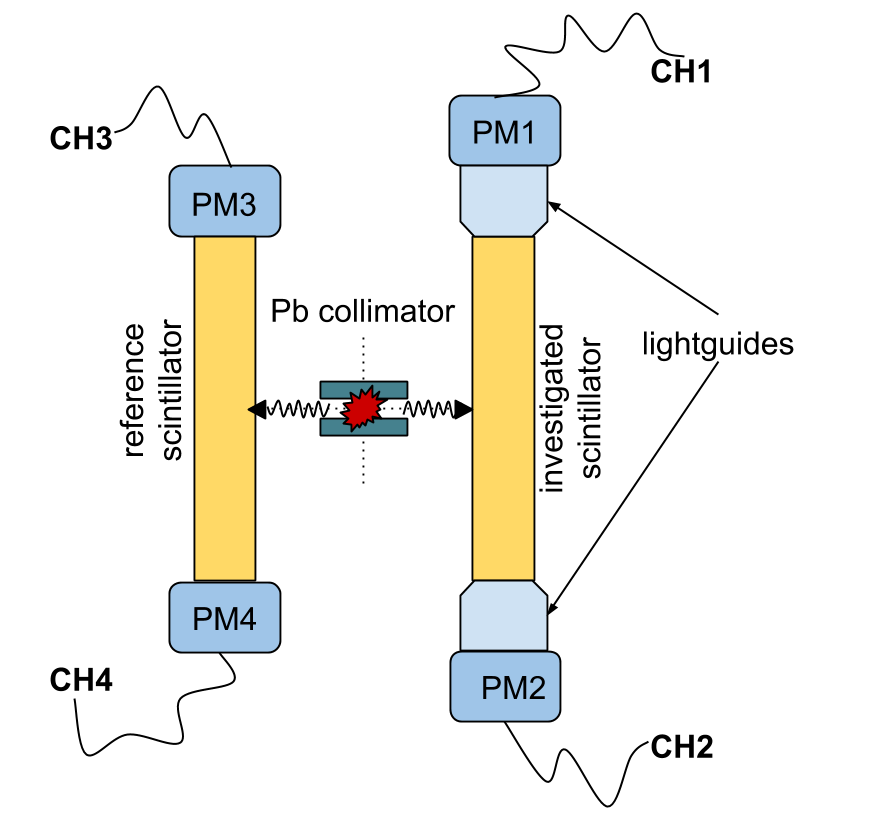}
\caption{\textit{The two strip \mbox{J-PET} setup simplified scheme. Scintillators during all measurements were wrapped in the vikuiti foil \cite{vikuiti}. Hamamatsu R4998 photomultipliers were used.}}\label{experimentalSetup}
\end{center}
\end{figure}

The measurements were performed with the two-strip \mbox{J-PET} system. The schematic arrangement of~the elements in the experimental setup is presented in the Figure~\ref{experimentalSetup}. In the one strip there are inserted lightguides (investigated strip) and the other one serves as~the reference. Sodium radioactive source is placed in the collimator with 1.5 mm slit in the middle between two strips, so that only the central part of~each scintillator is irradiated. This means that the expected time difference of~the signals registered on both ends of~the strip is zero. The measured time differences are the Gaussian-like distribution centred around zero. The parameter identified with the time resolution $ \sigma_t $ is the sigma of~this distribution. The idea of~the experiment is to~measure how the lightguides insertion decrease the precision of~time difference determination. Another test is to~estimate how much light is lost due to~their presence based on the position of~the Compton edge.

\begin{figure}[!h]
\begin{center}
\begin{subfigure}[C]{0.27\textwidth}
\includegraphics[width=\linewidth]{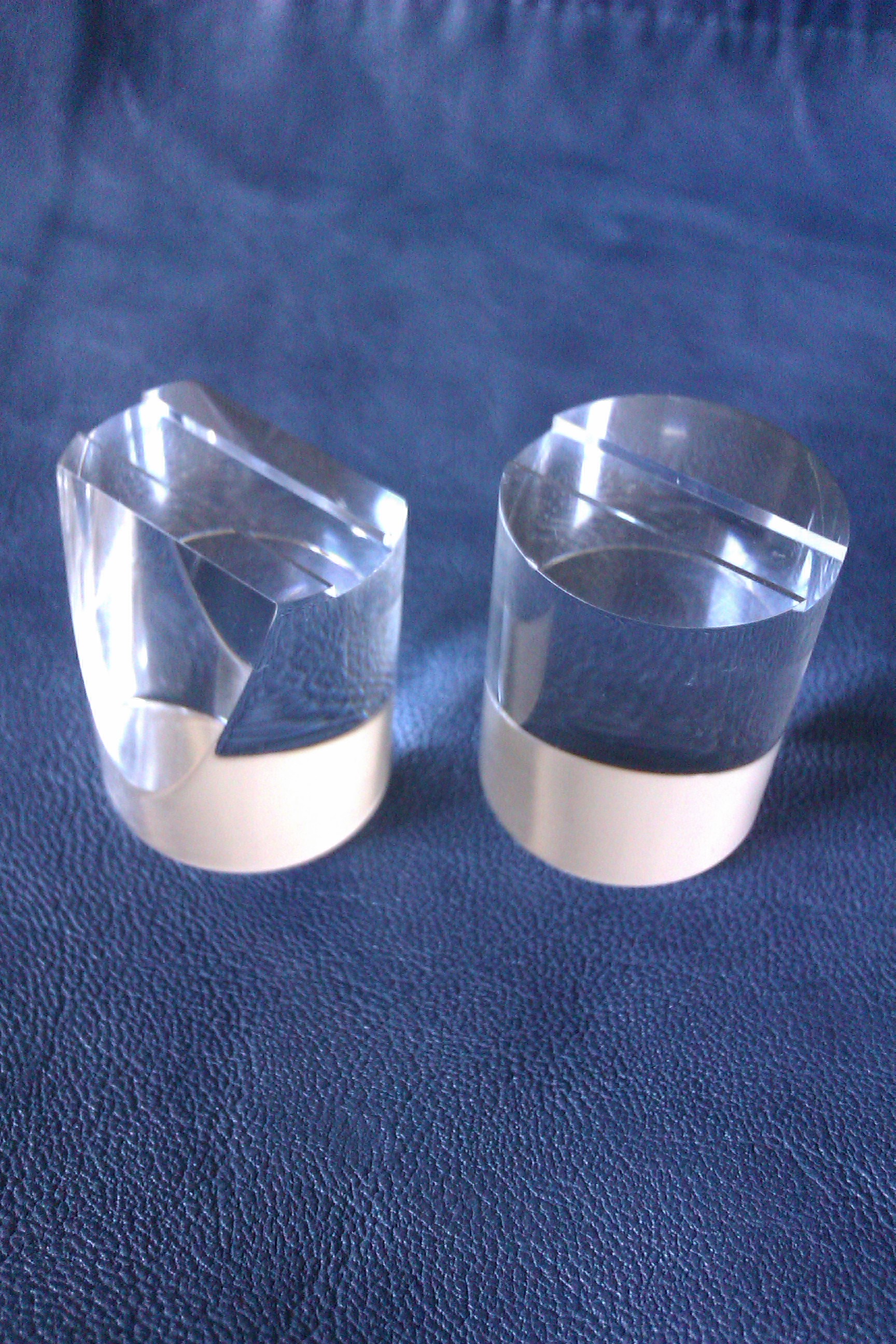}
\caption{}
\end{subfigure}
\begin{subfigure}[C]{0.27\textwidth}
\includegraphics[width=\linewidth]{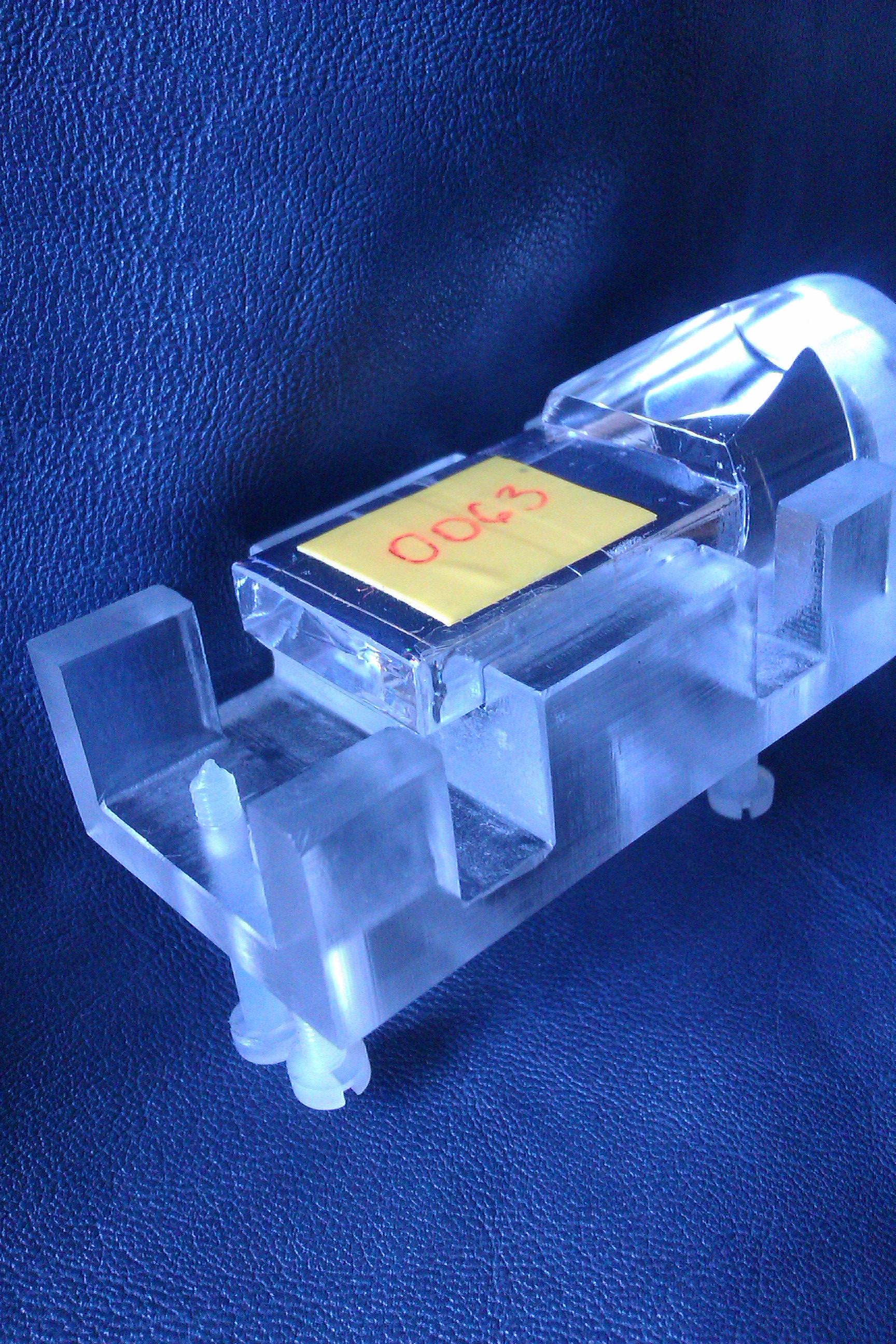}
\caption{}
\end{subfigure}
\begin{subfigure}[C]{0.27\textwidth}
\includegraphics[width=\linewidth]{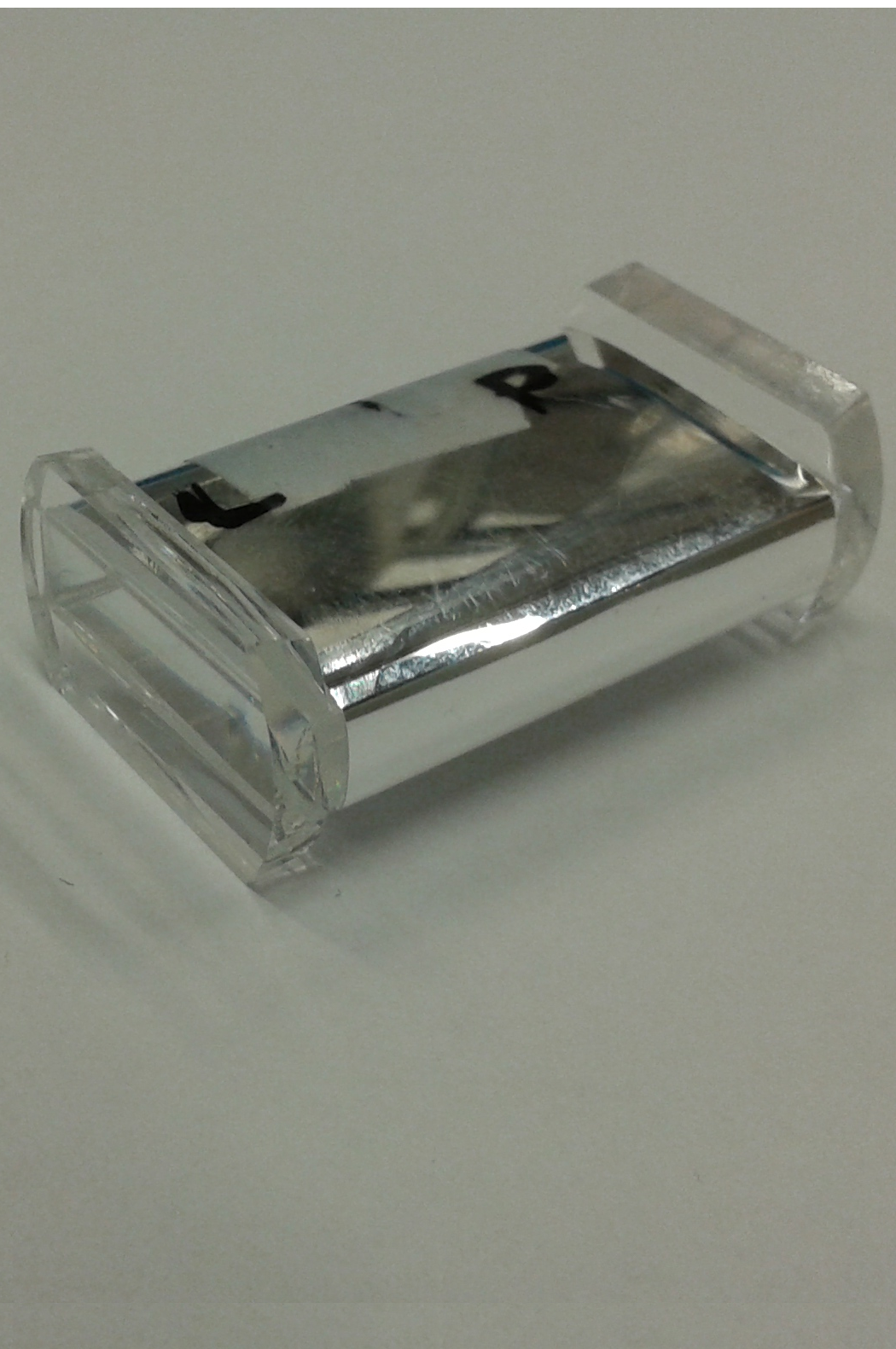}
\caption{}
\end{subfigure}
\end{center}
\vspace{-24pt}
\caption{(a) \textit{Trapezoidal and cylindrical lightguides with cuts.} (b) \textit{The pad used for stabilisation of~the~lightguides and scintillator connected with optical gel.} (c) \textit{Short lightguides glued to~the scintillator.}}\label{phtots}
\end{figure}

\begin{table}[!ht]
\centering
\vspace{11pt}
\begin{tabular}{| c | c | c |}
\hline
ELEMENT & TYPE & GEOMETRY \\
\hline
\begin{minipage}{0.3\textwidth}
\centering
scintillators \\
\end{minipage} & BC400 & \begin{minipage}{.4\textwidth}
\includegraphics[width=\linewidth]{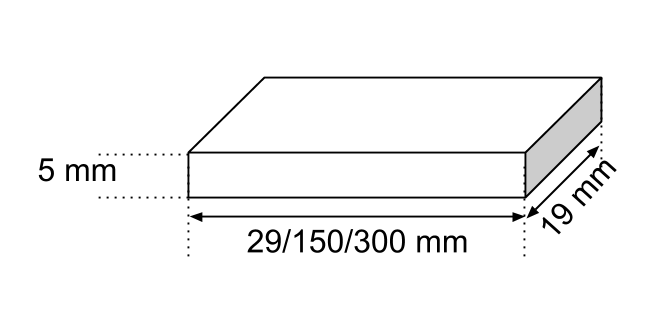}
\end{minipage} \\
\hline
\begin{minipage}{0.3\textwidth}
\centering
cylindrical \\
lightguides
\end{minipage} & BC800& 
\begin{minipage}{.4\textwidth}
\includegraphics[width=\linewidth]{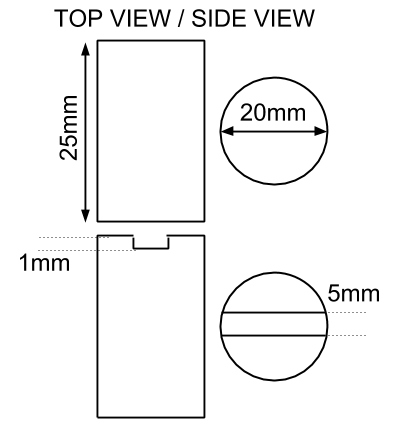}
\end{minipage} \\
\hline
\begin{minipage}{0.4\textwidth}
\centering
trapezoidal \\
lightguides
\end{minipage} & BC800 & 
\begin{minipage}{.45\textwidth}
\includegraphics[width=\linewidth]{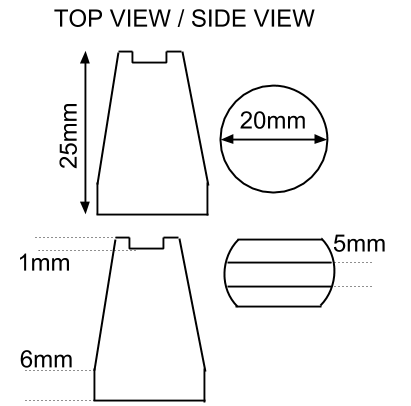}
\end{minipage} \\
\hline
\begin{minipage}{0.4\textwidth}
\centering
short \\
lightguides
\end{minipage} & BC800 & 
\begin{minipage}{.45\textwidth}
\includegraphics[width=\linewidth]{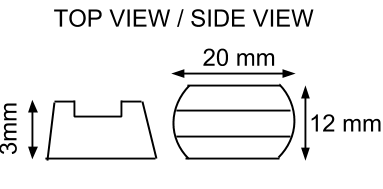}
\end{minipage} \\
\hline
\end{tabular}
\caption{\textit{Dimensions of~the scintillators and lightguides used in the experiment}.}\label{dimensions}
\end{table}
\FloatBarrier

The described measurement is possible only with the appropriate setup settings. The high voltage applied to~each photomultiplier was calculated so that the gain of~each photomultiplier was the same (Table \ref{voltages}). For more details about the photomultipliers characteristics and the gain calculation see \cite{raportTomka, artykulTomka}.

The oscilloscope trigger settings were chosen so that only the signals from the same annihilation were stored. This was provided by creating the known difference in time of~reaching the oscilloscope by the signals from two channels from different strips. It was done by inserting the signal cable of~the known time delay. Since the oscilloscope used allows only two channels to~be in the trigger conditions, channels 1 and 3 were used and an event was stored if both signals were higher than the given threshold within 60 ns window.
\begin{table}[!h]
\centering
\begin{tabular}{| c | c |}
\hline
PHOTOMULTIPLIER & HV [V]\\
\hline
PM1 & 1493\\
\hline
PM2 & 1500\\
\hline
PM3 & 1514\\
\hline
PM4 & 1497\\
\hline
\end{tabular}
\caption{\textit{High voltage settings for all performed measurements.}}\label{voltages}
\end{table}

\FloatBarrier
\section{Performed measurements} \label{PerformedMeasurements}

There were fifteen measurements performed in period from September 2014 to May~2015 (for details see Table \ref{measurements}). 

At first, six measurements were made with 3 cm scintillator in the investigated strip and with cylindrical lightguides, with trapezoidal lightguides (this measurement was done twice in order to~prove the repeatability of~the obtained results) and with both cylindrical and trapezoidal lightguides with the cuts matching the scintillator size. There was also done a~measurement with a~3 cm scintillator only to~determine its own time resolution which is considered as~a~reference. During these measurements in all cases the optical gel was used at~the connections.

Than, the most promising lightguides (trapezoidal with cuts) were chosen and permanently glued to~the 3 cm scintillator in order to~test the impact of~the optical glue.

As the short scintillators (3 cm) have very good time resolution, it was convenient to~use them to~clearly see the differences caused by the lightguides usage. But the measurements  were also done with longer scintillators (15 and 30 cm). It was tested whether the change in time resolution due to~the presence of~the lightguides is a~significant in comparison to~the time resolution of~the~longer strips.

Another part of~measurements was performed in March. The aim of~the two measurements with trapezoidal	lightguides was to~check the optical glue connection after a~few months. As~the results proved to~be repeatable, for the last measurement these lightguides were shortened up to~3 mm.

Finally in May 2015 the measurement with shortened lightguides was repeated in order to~confirm the final result.
\begin{landscape}
\begin{table}
\begin{small}
\centering
\begin{tabular}{|c|c|c|c|c|}
\hline
No. & lightguides  & lightguide scintillator & reference scintillator & date \\
	& type  		& length & length & \\
\hline
1 & cylindrical & 3 cm & 3 cm  & 08.09.2014 \\
\hline
2 & no lightguides & 3 cm & 3 cm & 11.09.2014 \\
\hline
3 & trapezoidal 1st test & 3 cm & 3 cm & 11.09.2014 \\
\hline
4 & trapezoidal 2nd test & 3 cm & 3 cm & 11.09.2014 \\
\hline
5 & cylindrical with cut & 3 cm & 3 cm  & 12.09.2014\\
\hline
6 & trapezoidal with cut & 3 cm  & 3 cm & 12.09.2014 \\
\hline
7 & no lightguides & 3 cm & 15 cm & 16.09.2014\\
\hline
8 & trapezoidal glued to~scintillator & 3 cm & 15 cm & 16.09.2014\\
\hline
9 & cylindrical cut & 3 cm & 15 cm & 17.09.2014\\
\hline
10 & cylindrical cut & 15 cm & 15 cm & 18.09.2014\\
\hline
11 & cylindrical cut & 30 cm & 15 cm & 19.09.2014\\
\hline  
12 & trapezoidal glued to~scintillator 1st test & 3 cm & 3 cm & 17.03.2015\\
\hline 
13 & trapezoidal glued to~scintillator 2nd test & 3 cm & 3 cm & 17.03.2015\\
\hline 
14 & shortened trapezoidal glued to~scintillator 1st test & 3 cm & 3 cm & 19.03.2015\\
\hline
15 & shortened trapezoidal glued to~scintillator 2nd test & 3 cm & 3 cm & 19.05.2015\\
\hline
\end{tabular}
\caption{\textit{Performed measurements in chronological order. "No lightguides" stands for the measurement with scintillators only.}}\label{measurements}
\end{small}
\end{table}
\end{landscape}

\FloatBarrier
\chapter{Data Analysis}
For each event amplitude of~four signals was sampled in the time domain with 100~ps step.

The first part of~the analysis is the selection of~signals  appropriate for further calculations. At~this point all events with at~least one signal smaller than some user-defined threshold are rejected. As~the expected range of~amplitudes of~signals is known from the previous studies \cite{average}, the threshold value chosen for data analysis in this work is -70 mV.

\begin{figure}[!h]
\begin{center}
\includegraphics[width = \linewidth]{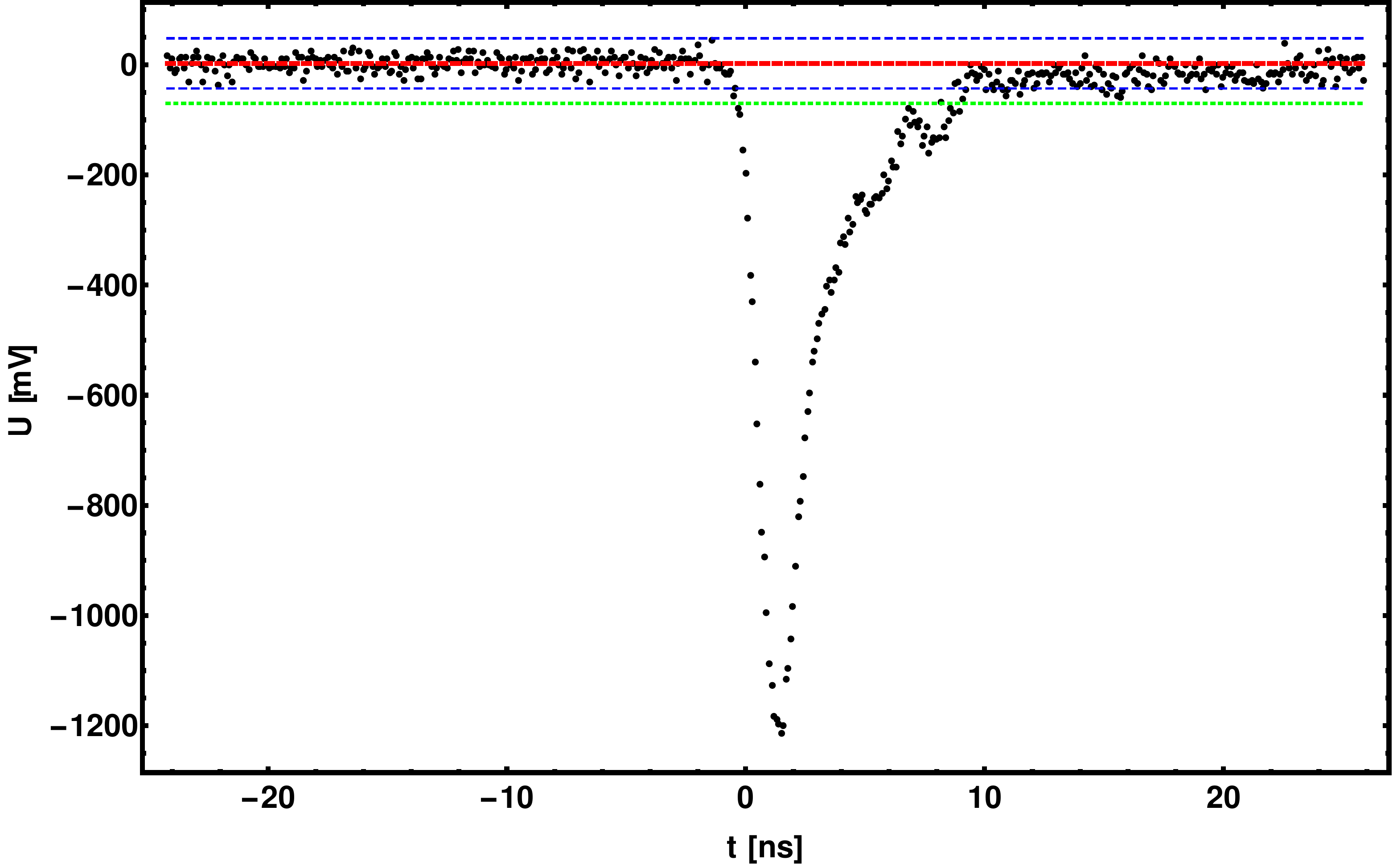}
\end{center}
\caption{\textit{Example signal registered by the oscilloscope. Green dotted line is the value of~the user-defined threshold. Red line is the mean value of~voltages of~first twenty points and blue lines limit the area of~noise. The offset value is calculated as~the mean value of~points from zero to~the first intersection of~the lower blue line and the signal. The offset level in this scale overlaps the mean value of~first twenty points.}}\label{signal}
\end{figure}

The next step is offset determination for each signal. The procedure begins with calculation of~the mean value and standard deviation of~amplitudes of~twenty first points in the signal. The noise level of~the signal is estimated as~the~mean value with three standard deviations (Figure~\ref{signal}).

Than, starting from the signal minimum, the voltage of~the left edge of~the signal are one by one compared with the noise range until the first point belonging to~noise is identified. The offset is calculated as~the mean value of~the points from zero to~that point. Once the offset is determined, the signal is shifted by its value~\cite{monika}.

The integral of~the shifted signal curve is the charge. Using the calibration for each photomultiplier it is possible to~calculate the charge of~the photoelectrons at~a~given high voltage value. The signal charge divided by the single photoelectron charge gives the number of~produced photoelectrons. This number is proportional to~the amount of~light that reached the photomultiplier. 

\begin{figure}
\begin{center}
\begin{subfigure}[c]{0.55\textwidth}
\includegraphics[width=\linewidth]{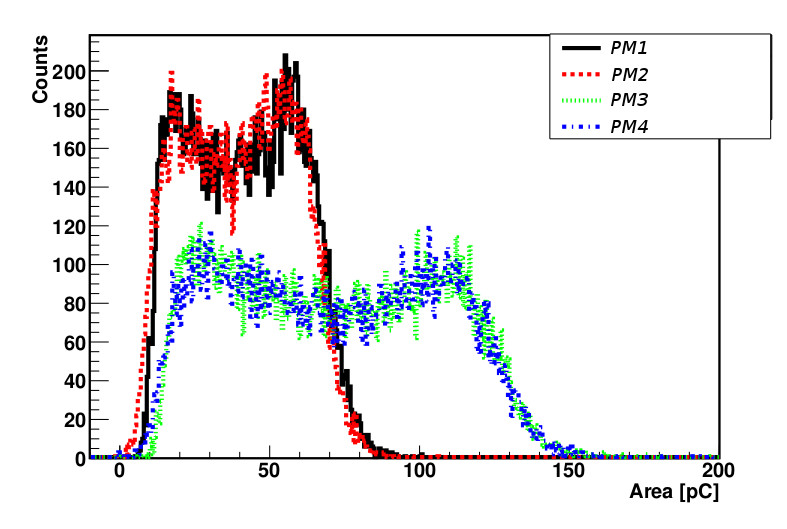}
\vspace{-26pt}
\caption{}\label{char_spect}
\end{subfigure}
\begin{subfigure}[c]{0.55\textwidth}
\includegraphics[width=\linewidth]{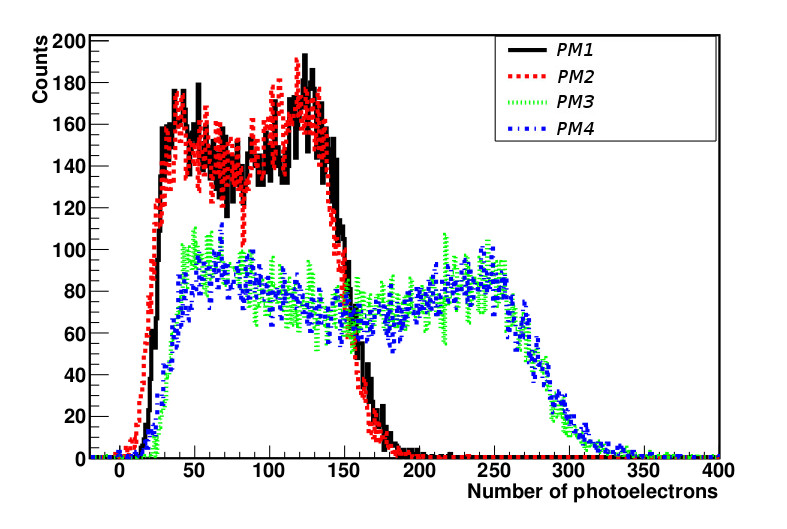}
\vspace{-26pt}
\caption{}\label{phe_spect}
\end{subfigure}
\begin{subfigure}[c]{0.55\textwidth}
\includegraphics[width=\linewidth]{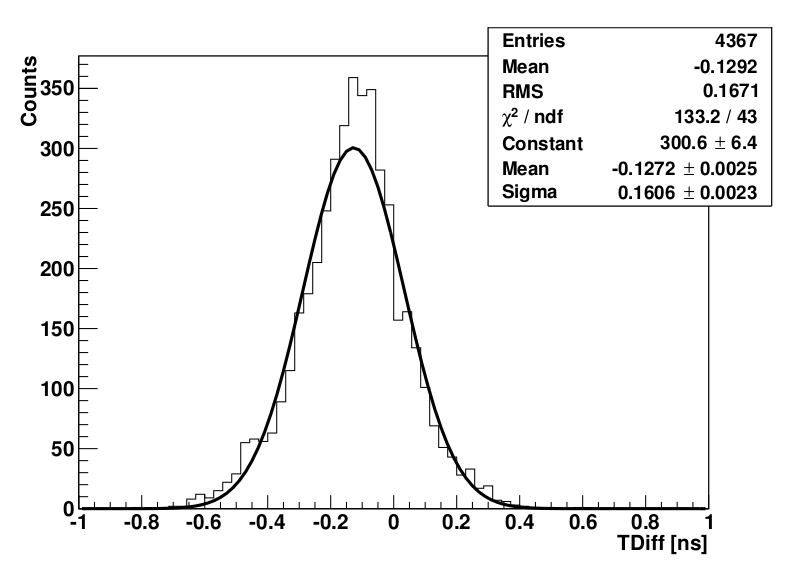}
\vspace{-26pt}
\caption{}\label{tdiff_spect}
\end{subfigure}
\end{center}
\caption{\textit{Example results (measurement no. 4: trapezoidal lightguides). (a) charge spectrum, (b) photoelectrons number spectrum. PMi stand for the respective photomultipliers. (c) Time difference spectrum between signals from PM1 and PM2. The non-zero position of~the peak is due to~different lengths of~the cables used in the setup.}}
\end{figure}

The described procedure is repeated for all signals from all channels. Example results are presented in the histograms \ref{char_spect} and \ref{phe_spect}. Compton edge is clearly visible. In order to~eliminate scattered events there is a~cut equal to~the mean value of~respective histogram made on the  photoelectrons number. Again, accepted events are those for which photoelectron number on all four channels are higher than the cut value. For more information about the Compton scattering impact on data analysis see \cite{szymon}.

The user-defined threshold is also shifted by the offset value. The time at~the new threshold is determined and it is used in further calculations. Particularly, the time difference between two signals is calculated based on this value. The time differences for all signals from one strip are computed and collected in the histogram. $ \sigma_t $ of~the distribution is obtained by fitting a~Gaussian function. An example histogram with fitted function is presented in the Figure~\ref{tdiff_spect}. For more information about the data analysis see also [22-24]. 

As the number of~photoelectrons is proportional to~the amount of~light reaching the photomultiplier, in order to~compare the light outputs obtained for tested lightguides, the spectra of~photoelectrons number were closely investigated. The value identified with the best light output is the number of~photoelectrons at~the center of~the Compton edge (where the center is defined as~the point in the half maximum of~the Compton edge).

Two functions were candidates to~fit the photoelectrons number spectra. The Fermi function given by the equation:
\begin{equation}
	f(x) = \frac{p_0}{e^{\frac{x - p_1}{p2}}+1} + p_3 \label{eq:fermi}
\end{equation}
where $ p_0 $ corresponds to~the amplitude, $ p_1 $ is a~horizontal translation, $ p_2 $ scales the edge of~the distribution, $ p_3 $ is a~vertical translation.
The advantage of this approach is the fact that the argument for which the function achieves its half maximum is obtained directly from the fit parameters (parameter $ p_1 $ in the equation \ref{eq:fermi}). a~considerable drawback is the necessity of~precise determination of~the fit beginning as~Fermi function is very sensitive to~this value.

The second tested function is Novosibirsk function \cite{novosybirsk}:
\begin{equation}
	F(x) = a~\exp\left(\frac{-\ln^2 \left(1 + \frac{\tau \left(x-x_0\right)}{\sigma}\frac{\sinh(\tau \sqrt{\ln 4})}{\tau \sqrt{\ln 4}}\right)}{ 2 \tau^2 } + \tau^2\right)\label{eq:nvs}
\end{equation}
where: $ a~$ - height of~the peak, $ x_0 $ - peak position, $ \sigma $ - width of~the peak, $ \tau $ - parameter responsible for the function decreasing.

The shape of~Novosibirsk function eliminates the problem of~the fit beginning providing a~stable method of~determining Compton edge center. The Compton edge center position ($ x_c $) was calculated for the condition: $ F(x_c) = \frac{1}{2}A $.

\begin{figure}[!h]
\begin{center}
\begin{subfigure}[c]{0.49\textwidth}
\includegraphics[width=\linewidth]{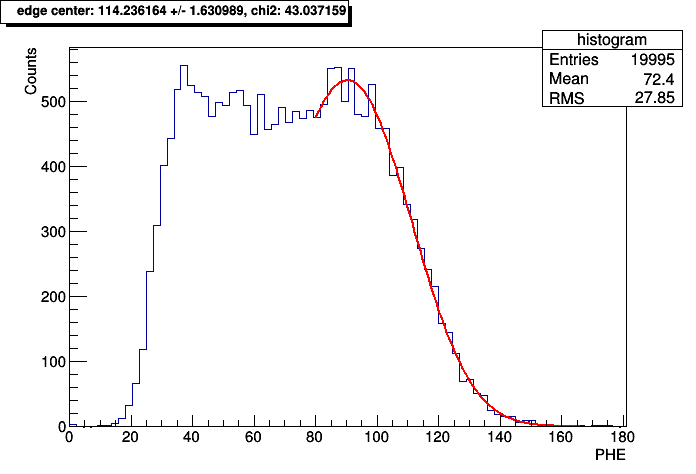}
\caption{}
\end{subfigure}
\begin{subfigure}[c]{0.49\textwidth}
\includegraphics[width=\linewidth]{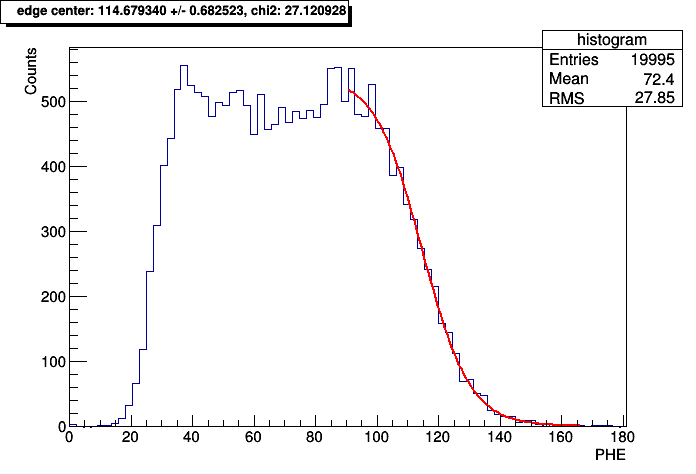}
\caption{}\label{sampleFermi}
\end{subfigure}
\end{center}
\caption{\textit{Example results of (a) Novosibirsk and (b) Fermi function fit to~the photoelectrons number spectra (signals from PM1, measurement no. 4: trapezoidal lightguides).}}\label{sampleNvs}
\end{figure}

Once the maximum is obtained it is possible to~fit the Fermi function with the beginning of~the fit range set as~the $ x $ coordinate of~the maximum value of~the Novosibirsk function. Sample fits obtained by these two methods are presented in the Figure~\ref{sampleNvs}.

Determination of the center point of the Compton edge gives the same result within error limits for both functions. However, the Novosybirsk function describes the distribution in the wider range, therefore this function was used for further analysis.

\FloatBarrier
\chapter{Results and Conclusions}
\begin{figure}
\begin{center}
\includegraphics[width=\linewidth]{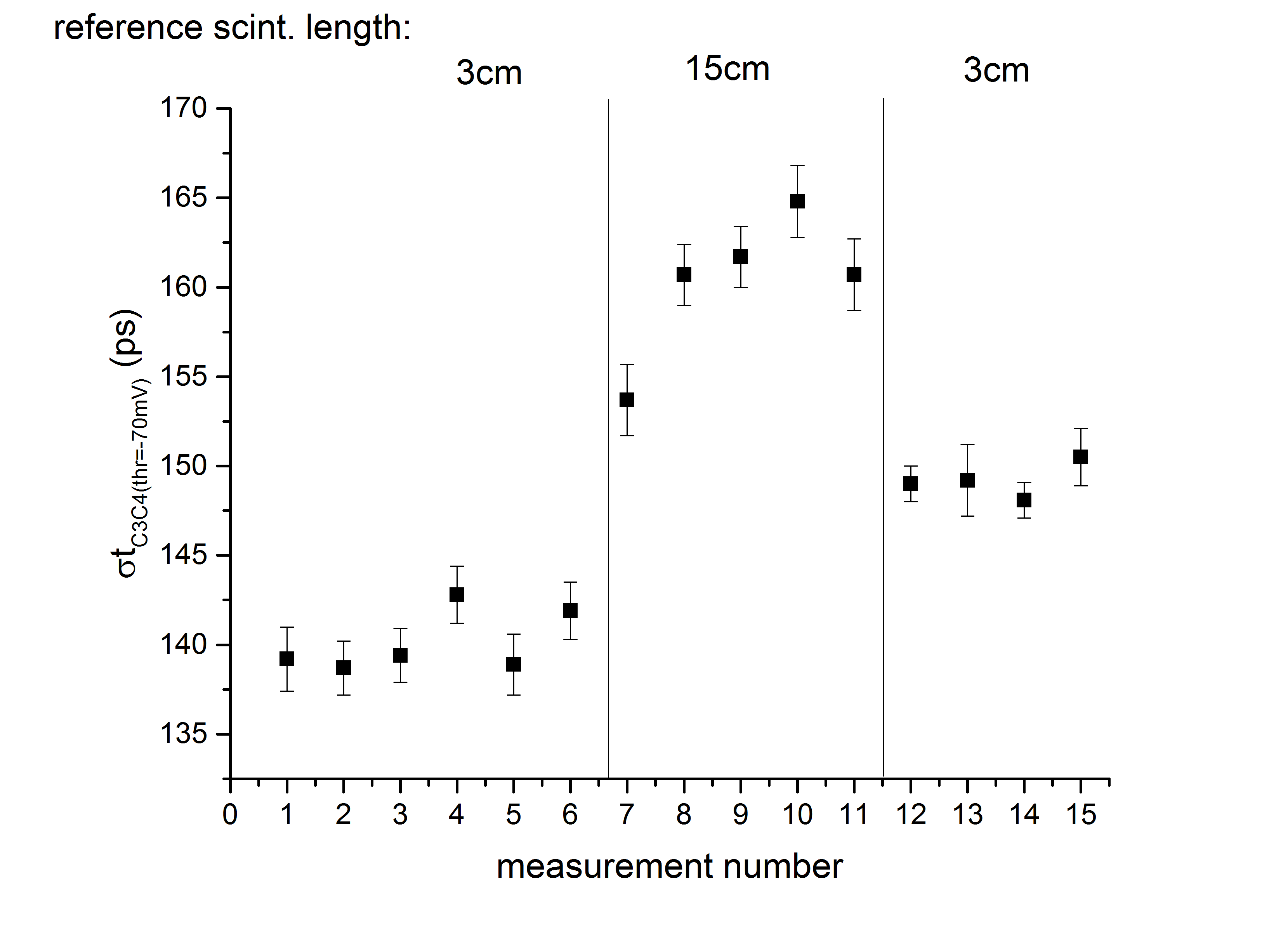}
\end{center}
\caption{\textit{Time resolution of the reference strip for the measurements: 1  - cylindrical lightguides,  2  - no lightguides,  3  - trapezoidal lightguides 1st test,  4  - trapezoidal lightguides 2nd test,  5  - cylindrical lightguides with cuts,  6  - trapezoidal lightguides with cuts,  7  - no lightguides,  8  - trapezoidal lightguides glued to scintillator,  9  - cylindrical lightguides with cuts and 3 cm scintillator,  10  - cylindrical lightguides with cuts 15 cm scintillator,  11  - cylindrical lightguides with cuts 30 cm scintillator,   12  - trapezoidal lightguides glued to scintillator 1st. test,  13  - trapezoidal lightguides glued to scintillator 2nd test, 14 - shortened trapezoidal lightguides glued to the scintillator 1st test, 15 - shortened trapezoidal lightguides glued to the scintillator 2nd test.}}\label{fig:reference}
\end{figure}

The time resolution for the reference strip is presented in the Figure~\ref{fig:reference} (see also Table \ref{measurements}). It is clearly visible that the time resolution for the reference strip is constant within error bars independently from the lightguides measurement in the second strip. On the other hand, as~expected, this time resolution depends on the length of~the scintillating strips. The 10 ps worse result for the time resolution for the measurements 12-15 (with respect to~measurements 1-6) has not been explained however it is negligible with respect to the effects observed at~the investigated strip. In addition, the increased length of~the scintillating strip from 3~cm to~15~cm increase the time resolution by $\sim$20 ps.

\begin{figure}
\begin{center}
\includegraphics[width=\linewidth]{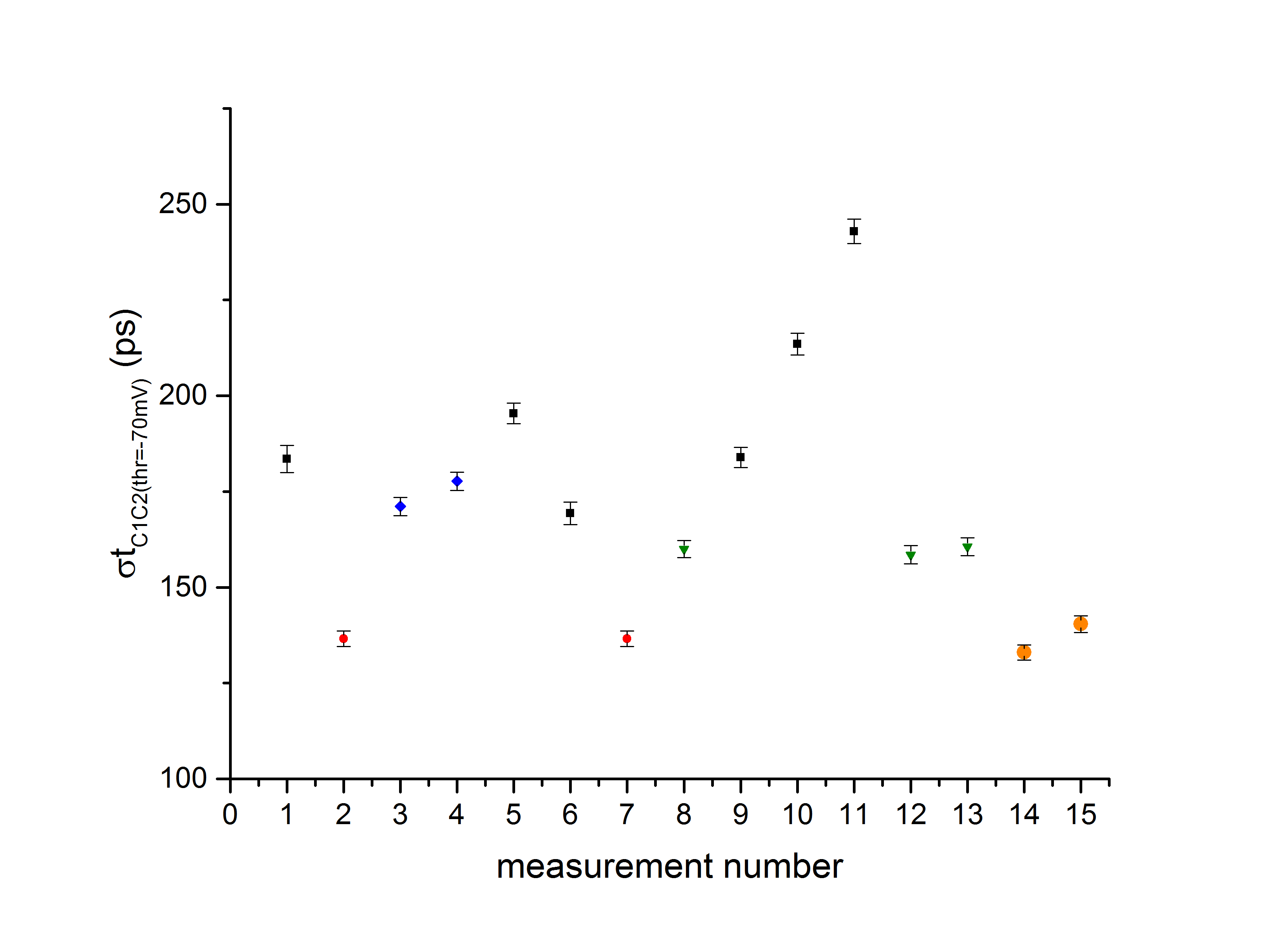}
\end{center}
\caption{\textit{Time resolution of the reference strip for the measurements: 1  - cylindrical lightguides,  2  - no lightguides,  3  - trapezoidal lightguides 1st test,  4  - trapezoidal lightguides 2nd test,  5  - cylindrical lightguides with cuts,  6  - trapezoidal lightguides with cuts,  7  - no lightguides,  8  - trapezoidal lightguides glued to scintillator,  9  - cylindrical lightguides with cuts and 3 cm scintillator,  10  - cylindrical lightguides with cuts 15 cm scintillator,  11  - cylindrical lightguides with cuts 30 cm scintillator,   12  - trapezoidal lightguides glued to scintillator 1st. test,  13  - trapezoidal lightguides glued to scintillator 2nd test, 14 - shortened trapezoidal lightguides glued to the scintillator 1st test, 15 - shortened trapezoidal lightguides glued to the scintillator 2nd test. Red, blue, green and orange data points correspond to the measurements with the same conditions.}}\label{fig:lightguide}
\end{figure}

Results of~the measurements with lightguides are presented in the Figure~\ref{fig:lightguide}.  Measurements 1 and 3 for cylindrical and trapezoidal lightguides respectively show that the latter has better time resolution which is also proved in the measurements with cut lightguides (measurements 5 and 6). On the other hand there is no difference observed between the results for lightguides of~the same shape with and without the cuts. The small improvement is visible with the optical glue used instead of~gel (measurement 3 and 8, 12, 13 respectively). Results 9-11 show that each $\sim$15 cm of~additional scintillator length adds $\sim$30 ps in the time resolution while application of~3 cm lightguide adds $ \sim $20 ps in $ \sigma_t $ (comparison of~the results 2, 7 and 1, 3-6). Finally, the time resolution for the 3 mm lightguides is comparable with the results without any lightguides. Therefore this element is suggested to be used in the full barrel \mbox{\mbox{J-PET}} detector.

\begin{figure}
\begin{center}
\includegraphics[width=\linewidth]{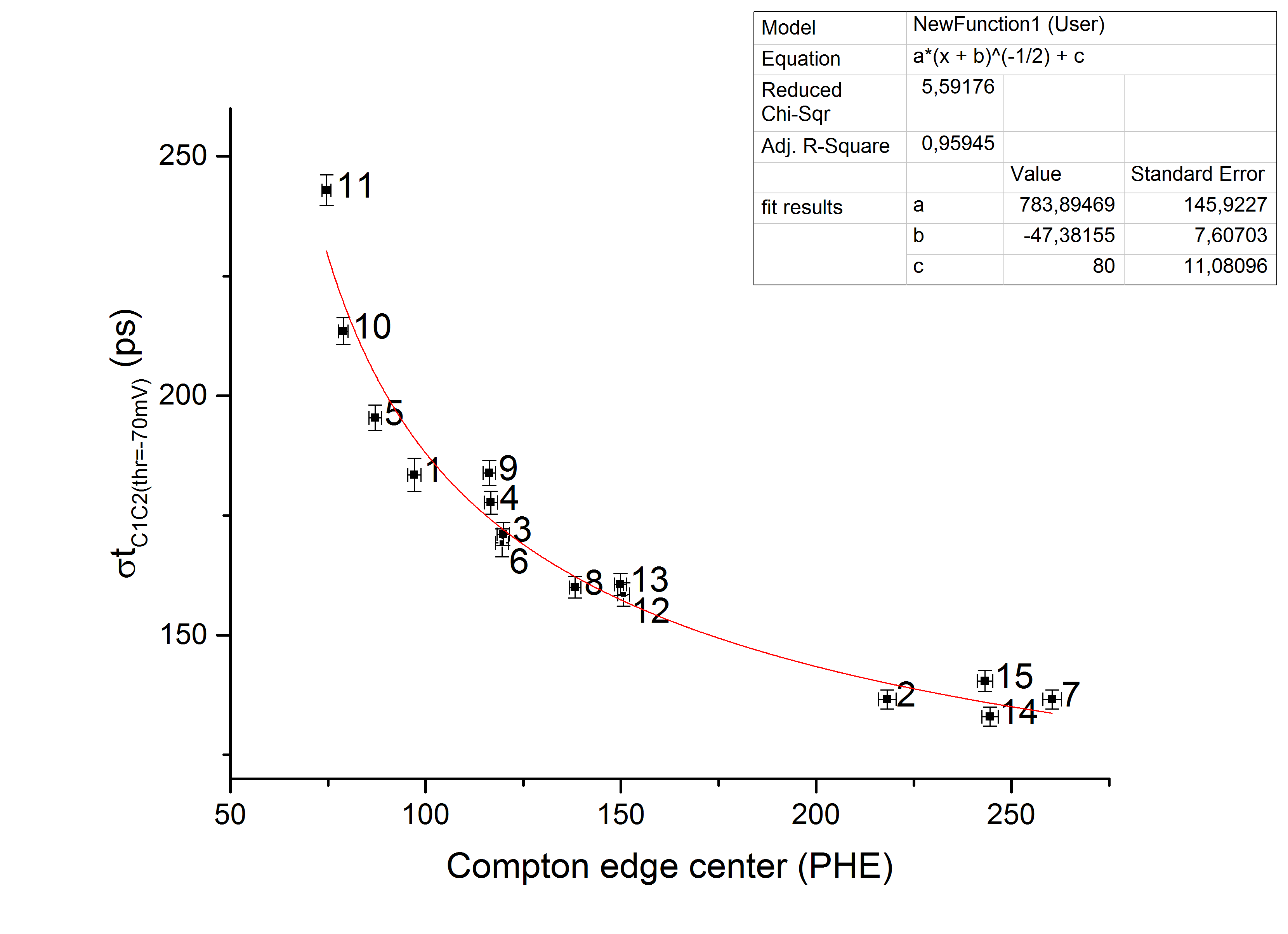}
\end{center}
\caption{\textit{Time resolution as a function of the number of photoelectrons. Red function \mbox{$f(x) = a (x + b ) ^{-0.5} + c $} corresponds to the fit to data points: 1  - cylindrical lightguides,  2  - no lightguides,  3  - trapezoidal lightguides 1st test,  4  - trapezoidal lightguides 2nd test,  5  - cylindrical lightguides with cuts,  6  - trapezoidal lightguides with cuts,  7  - no lightguides,  8  - trapezoidal lightguides glued to scintillator,  9  - cylindrical lightguides with cuts and 3 cm scintillator,  10  - cylindrical lightguides with cuts 15 cm scintillator,  11  - cylindrical lightguides with cuts 30 cm scintillator,   12  - trapezoidal lightguides glued to scintillator 1st. test,  13  - trapezoidal lightguides glued to scintillator 2nd test, 14 - shortened trapezoidal lightguides glued to the scintillator 1st test, 15 - shortened trapezoidal lightguides glued to the scintillator 2nd test.}}\label{sigmavsphe}
\end{figure}

In addition the dependence of the signal resolution on the statistical fluctuations of number of photoelectrons (PHE) in the form of $ \sigma_t \sim $PHE$^{-0.5}$ was checked (\ref{sigmavsphe}). 

\newpage
\null
\vspace{60pt}
\begin{center}
{\Large \textbf{\textit{Acknowledgements}}}
\end{center}
\vspace{24pt}

\selectlanguage{polish}
\textit{I would like to express my deepest gratitude to my supervisor, Dr Eryk Czerwiński for his effort, valuable suggestions and guidance from the very beginning of my experimental work.}
\vspace{15pt}

\textit{I am also grateful to Prof. Paweł Moskal for all his remarks and the opportunity to work in the \mbox{\mbox{J-PET}} group. }
\vspace{15pt}

\textit{I also thank Ewelina Kubicz, Magdalena Skurzok, Anna Wieczorek and Szymon Niedźwiecki for their advice and patient help even with the simplest things. Each of them contributed something to this work.}
\vspace{15pt}

\textit{I wish to acknowledge financial support by the Polish National Center for Development and Research through grant INNOTECH-K1/IN1/64/159174/NCBR/1.}
\newpage
\null
\selectlanguage{english}

\end{document}